\documentclass{ieeetj}
\usepackage{cite}
\usepackage{amsmath,amssymb,amsfonts}
\usepackage{algorithmic}
\usepackage{hyperref}
\hypersetup{hidelinks=true}
\usepackage{algorithm,algorithmic}
\usepackage{graphicx}
\usepackage{textcomp}
\usepackage{wrapfig,colortbl}
\usepackage{dblfloatfix}
\usepackage{tabularx}
\usepackage[table]{xcolor}
\usepackage{subcaption}
\let\labelindent\relax
\usepackage{enumitem}

\newcommand{\figurePath}{figures}

\def\BibTeX{{\rm B\kern-.05em{\sc i\kern-.025em b}\kern-.08em
    T\kern-.1667em\lower.7ex\hbox{E}\kern-.125emX}}
\AtBeginDocument{\definecolor{tmlcncolor}{cmyk}{0.93,0.59,0.15,0.02}\definecolor{NavyBlue}{RGB}{0,86,125}}

\def\OJlogo{\vspace{-4pt}$<$Society logo(s) and publication title will appear here.$>$}
\def\seclogo{\vspace{10pt}$<$Society logo(s) and publication title will appear here.$>$}

\def\authorrefmark#1{\ensuremath{^{\textbf{#1}}}}

\begin{document}

\receiveddate{XX Month, XXXX}
\reviseddate{XX Month, XXXX}
\accepteddate{XX Month, XXXX}
\publisheddate{XX Month, XXXX}
\currentdate{XX Month, XXXX}
\doiinfo{XXXX.2025.1234567}

\markboth{}{Dahunsi {et al.}}

\title{Orthogonal Plane-Wave Transmit-Receive Isotropic-Focusing Micro-Ultrasound (OPTIMUS) with Bias-Switchable Row-Column Arrays }
\author{Darren Dahunsi\authorrefmark{1}, \IEEEmembership{Student Member, IEEE}, 
Randy Palamar\authorrefmark{1}, \IEEEmembership{Student Member, IEEE},
Tyler Henry\authorrefmark{1}, \IEEEmembership{Student Member, IEEE}, 
Mohammad Rahim Sobhani\authorrefmark{1,2}, \IEEEmembership{Student Member, IEEE}, 
Negar Majidi\authorrefmark{1}, \IEEEmembership{Student Member, IEEE}, 
Joy Wang\authorrefmark{1}, \IEEEmembership{Student Member, IEEE}, 
Afshin Kashani Ilkhechi\authorrefmark{1,2}, \IEEEmembership{Student Member, IEEE}, 
 and 
Roger Zemp\authorrefmark{1,2}, \IEEEmembership{Member, IEEE}
}
\affil{Department of Electrical and Computer Engineering, University of Alberta, Edmonton, AB T6G 2R3, Canada}
\affil{CliniSonix Inc., Edmonton, AB  T5J 4P6, Canada}
\corresp{Corresponding author: Darren Dahunsi (email: dahunsi@ualberta.ca).}
\authornote{RJZ and MRS are directors and shareholders of CliniSonix Inc., which provided partial support for this work. RJZ is a founder 
and director of OptoBiomeDx Inc., which, however, did not support this work. RJZ is also a founder and shareholder of IllumiSonics Inc.,which, however, did 
not support this work.
    We gratefully acknowledge funding from the National Institutes of Health  (1R21HL161626-01 and EITTSCA R21EYO33078), Alberta Innovates (AICE 202102269, 
CASBE 212200391 and LevMax 232403439), MITACS (IT46044 and IT41795), CliniSonix Inc., NSERC (2025-05274) and a NSERC CGS-D to DD (588860 - 2024), the Alberta Cancer Foundation and the Mary 
Johnston Family Melanoma Grant (ACF JFMRP 27587), the Government of Alberta Cancer Research for Screening and Prevention Fund (CRSPPF 017061), an Innovation 
Catalyst Grant to MRS, INOVAIT (2023-6359), and an IBET Momentum Fellowship to DD. We are grateful to the nanoFAB staff at the University of Alberta for 
facilitating array fabrication.}

\begin{abstract}
      High quality structural volumetric imaging is a challenging goal to achieve with modern ultrasound transducers. Matrix probes have limited fields of view 
      and element counts, whereas row-column arrays (RCAs) provide insufficient focusing. In contrast, Top-Orthogonal-to-Bottom-Electrode (TOBE) arrays, also 
      known as bias-switchable RCAs can enable isotropic focusing on par with ideal matrix probes, with a field of view surpassing conventional RCAs. Orthogonal 
    Plane-Wave Transmit-Receive Isotropic-Focusing Micro-Ultrasound (OPTIMUS) is a novel imaging scheme that can use TOBE arrays to achieve nearly isotropic focusing
    throughout an expansive volume. This approach extends upon a similar volumetric imaging scheme, Hadamard Encoded Row Column Ultrasonic Expansive 
    Scanning (HERCULES), that is even able to image beyond the shadow of the aperture, much like typical 2D matrix probes. We simulate a grid of scatterers to 
    evaluate how the resolution varies across the volume, and validate these simulations experimentally using a commercial calibration phantom. Experimental 
      measurements were done with a custom fabricated TOBE array, custom biasing electronics, and a research ultrasound system. Finally we performed ex-vivo 
      imaging to assess our ability to discern structural tissue information.
\end{abstract}


\maketitle

\section{Introduction}

\begin{figure*}
    \centerline{\includegraphics[width=\linewidth]{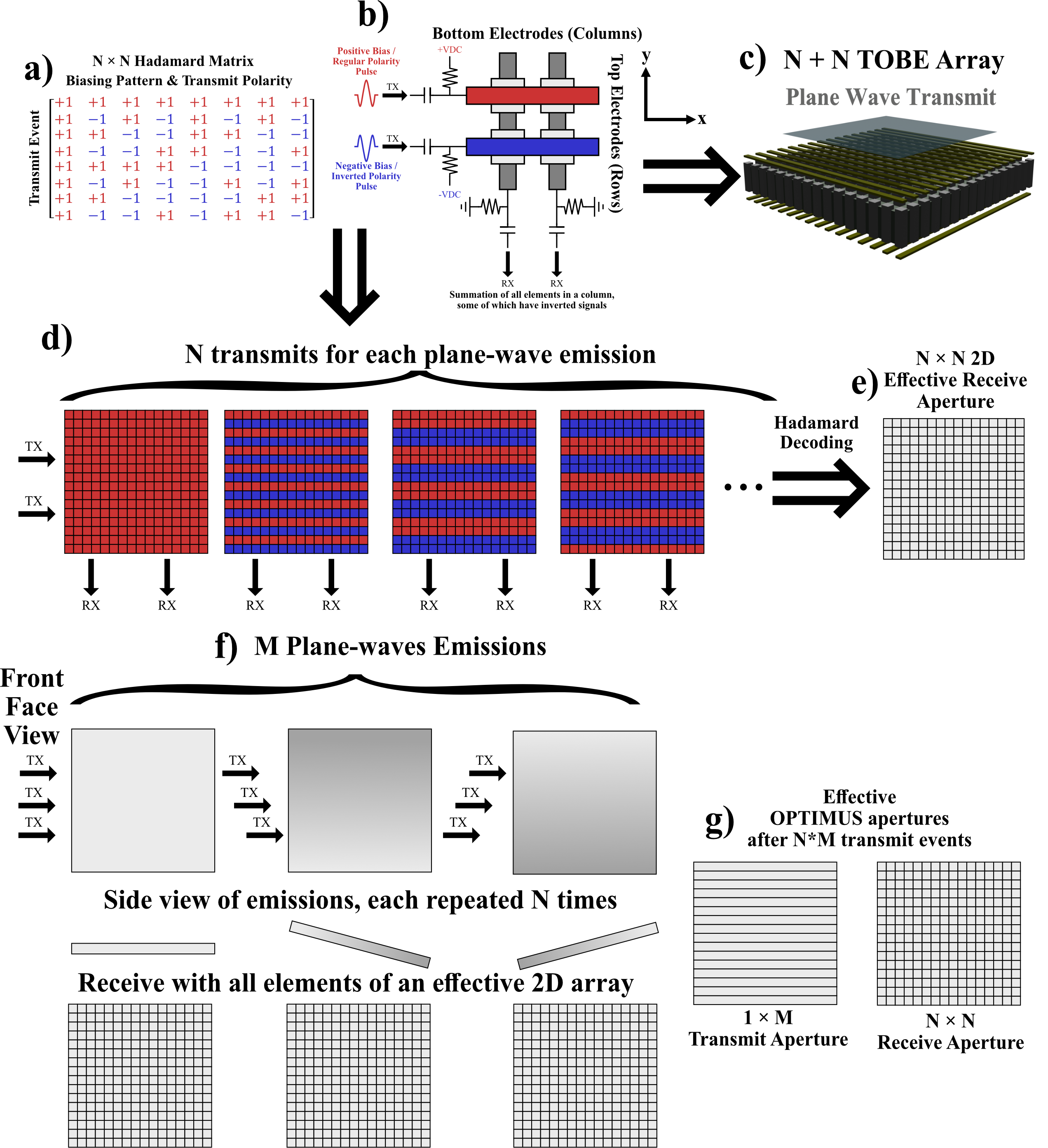}}
    \caption{Summary of the OPTIMUS imaging scheme for an arbitrarily sized array of N rows and N columns with N$\times$M transmit events. In a) we use the rows of a Hadamard matrix to 
    determine the biasing and transmit polarities that are applied on the rows as shown in b), with each column of the matrix corresponding to a different row of the array. 
    c) shows a render of the effective layout of the element electrodes on the transducer, and an effective transmit waveform. In d) we can see the effective 
    bias pattern between each transmit event. For a set of N transmits the transmitted waveform will not change, but the received waveform will, and decoding will 
    result in the receive aperture shown in  e). f) We will repeat those N transmits for M different plane-wave angles and obtain the effective imaging aperture 
    shown in g).}
    \label{fig:imagingScheme}
\end{figure*}

\label{sec:introduction}\IEEEPARstart{V}{olumetric}, ultrasound imaging is an area of growing interest as transducer and computational capabilities increase and 
realtime volumetric imaging becomes more feasible. Newer transducer architectures such as matrix probes and row column arrays have potential to enable 
clinical applications that require accurate 3-dimensional views of physiology. Volumetric views also better enable certain advanced imaging modes like 
elastography and blood flow imaging. Furthermore, 3D imaging may mitigate operator dependence.
\cite{Smith:1991:HighSpeedSystemA, Welch:2001:Freehand3D, Austeng:2002:Sparse2DArray3DImaging, Yen:2004:3DRectilinearMultiplex, Chang:2015:GraphSegmentation3D, 
Downey:1995:3DVascularDoppler, Huang:2015:3DLinearStrain}

\begin{table}[t]
    \caption{Imaging Parameters}
    \label{table:simDieParams}
    \setlength{\tabcolsep}{3pt}
    \begin{tabular}{p{150pt}p{75pt}}
        \hline\hline
        a) Transducer Parameters           &
        Value                           \\
        \hline
        2D array size                   &
        128 $+$ 128                     \\
        Transmit Count                  &
        128                             \\
        Pitch                           &
        250 $\mu$m                      \\
        Kerf                            &
        30 $\mu$m                       \\
        Center frequency                &
        6.25 MHz                        \\
        \hline\hline
        b) Simulation Parameters           &
        Value                           \\
        \hline
        Number of excitation cycles     &
        1                               \\
        Speed of Sound                  &
        1540 m/s                        \\
        Sampling frequency              &
        50 MHz                          \\
        \hline\hline
        c) Experimental Parameters           &
        Value                           \\
        \hline
        Chirp Length     &
        20e-6 $\mu$s                               \\
        Chirp Bandwidth     &
        4$-$8 MHz                               \\
        Speed of Sound                  &
        1452 m/s                        \\
        Sampling frequency              &
        16.6 MHz                          \\
        \hline\hline
        d) Imaging Parameters &
        Value                           \\
        \hline
        Transmit Count (OPTIMUS) &
        1152 \\
        Transmit Count (HERCULES, VLS, TPW) &
        128 \\
        Angle Count (OPTIMUS) &
        9 \\
        Maximum Angle (OPTIMUS) &
        $\pm10^\circ$ \\
        Maximum Angle (TPW) &
        $\pm10^\circ$ \\
        \hline\hline
    \end{tabular}
\end{table}

Fully wired matrix probes have been proposed to achieve volumetric images, but the B-scan image quality is often inferior to clinical linear arrays which have
a wide footprint and thus a low f-number. Scaling matrix probes up to similar footprints as these linear arrays is non-trivial. As the channel count of a 2D array 
increases quadratically, existing matrix probes are limited to small footprints, with 32$\times$32 matrix probes being the largest commercial probes currently 
available. X-Matrix probes with microbeamformers require non-ideal beamforming approximations and wide beam transmits which 
negatively impact image quality. \cite{Smith:1991:HighSpeedSystemA, vonRamm:1991:HighSpeedSystemB, Heiles:2019:LocalizationMicrosopyMatrix32, 
Zhang:2021:CEUS_Sparse_Matrix}

Conversely, Row Column Arrays (RCAs) are able to have large footprints and large element counts, which promise high numerical apertures. However, because
they are effectively two orthogonal linear arrays, they suffer from a lack of two-way focusing and are also constrained to imaging directly below the shadow of 
the aperture. There are previous RCA imaging schemes that use this architecture to perform volumetric imaging. Virtual line-source imaging (VLS) uses an array of virtual line sources as the transmit aperture, while Tilted Plane-Wave (TPW) imaging uses a 
series of tilted plane-waves \cite{Awad:2009:3DSpatialCompoundRCA, Flesch:2017:4DUltrafastRCA,Jørgensen:2023:RCABeamformer}. Combined with using the orthogonal array 
as a receive aperture, RCAs can only perform one-way cylindrical focusing on transmit or receive, but not both. This is unlike an ideal fully-wired matrix probe that can 
perform two-way spherical focusing, and this limitation causes reduced image quality.

Many modern volumetric imaging setups attempt to leverage the principles of ultrafast ultrasound imaging. Some of the primary benefits of ultrafast imaging are 
the decoupling of volume acquisition rate from volume resolution and the increased temporal coherence across the volume. On the other hand, insonifying the 
entire imaged region simultaneously reduces contrast \cite{Montaldo:2009:CoherentPW, Couade:2009:UltrafastHeart,Tanter:2014:UltrafastReview}. Extended transmits like chirp excitations can be used to mitigate these types of issues, but in turn 
potentially introduce problems like a high mechanical index and probe heating \cite{Kollmann:2004:SurfaceHeating, Calvert:2007:TransvaginalHeating, 
Nowicki:2020:ThermalMechanicalReview}. Thankfully, the increased surface area of large specialized RCA arrays can mitigate these issues.

Electrostrictive top-orthogonal-to-bottom-electrode (TOBE) arrays are bias-sensitive RCAs that allow new capabilities beyond what conventional piezoelectric RCAs 
can enable. These electrostrictives are not inherently piezoelectric until a bias voltage is applied. This bias-sensitivity means that we can selectively enable 
sections of a transducer array. When this property is applied to a row-column array this allows us to manipulate or encode the effective aperture of the 
transducer on transmit or receive. By selectively activating parts of the array we can conduct full synthetic aperture imaging akin to a fully wired 2D array,
however, due to the partial activation of the array, there is a significant decrease in SNR. Hadamard encoding presents a way to attain this fully wired capability without
sacrificing signal strength.

\begin{figure}
    \centering
    \includegraphics[width=0.9\columnwidth]{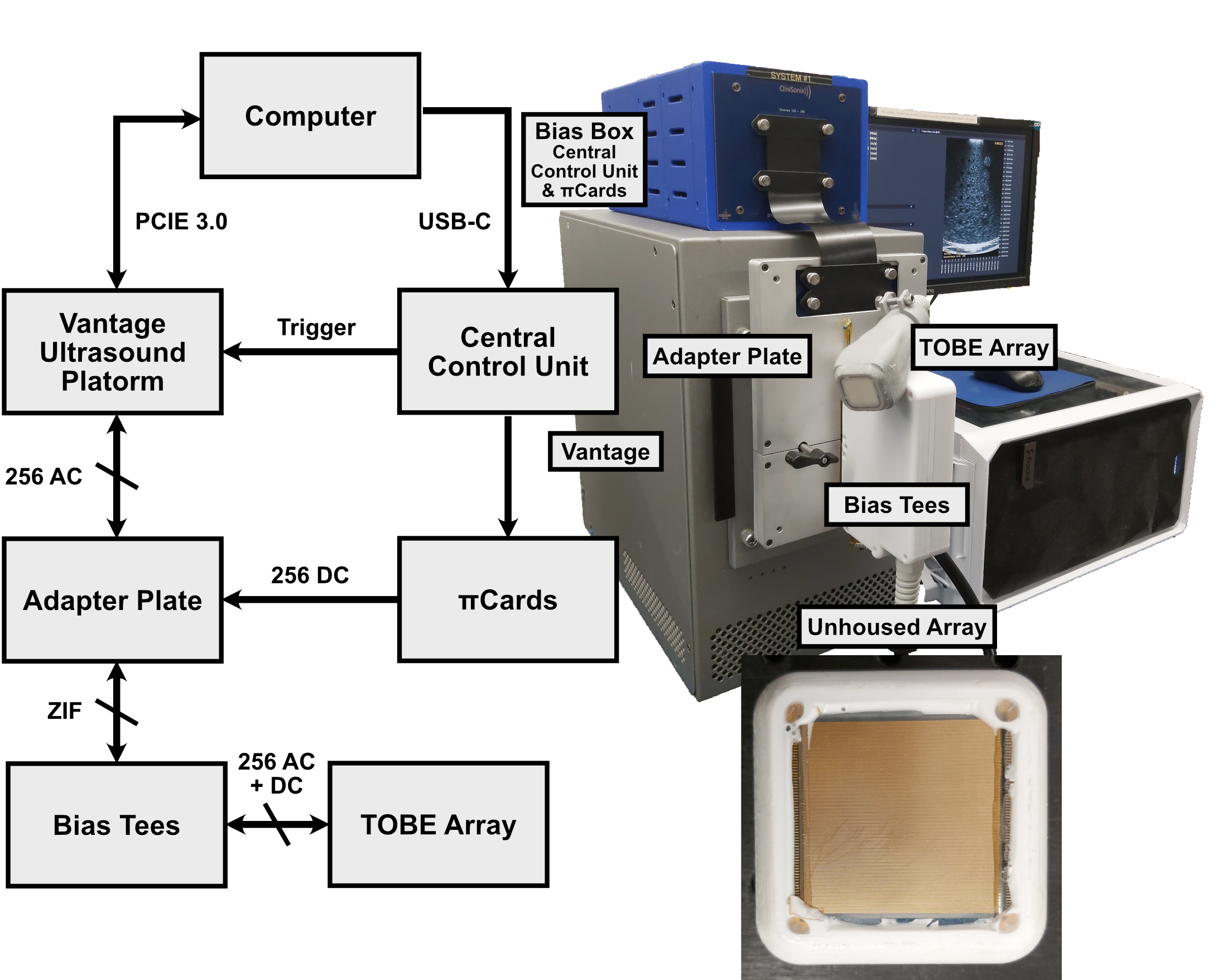}
    \caption{Setup for experimental imaging. We image with a Vantage Ultrasound Research platform, controlled by a host computer that also coordinates with the 
    Central Control Unit of custom biasing electronics. The Central Control Unit controls the high-voltage biasing cards ($\pi$Cards), while waiting on triggers 
    from the Vantage Unit to determine when to change the biasing pattern. The DC signals from the $\pi$Cards are mixed with AC signals from the Vantage and 
    coupled on a bias tee and delivered to the TOBE array. An unhoused array is also shown.}
    \label{fig:experimentalSetup}
\end{figure}

In previous work we have shown that we can leverage hadamard encoding on bias-sensitive row column arrays to perform ultrafast imaging with a 
variety of imaging methods. In one method we perform transmit encoding and image 2D planes with high resolution by leveraging the large numerical aperture of 
our arrays \cite{Ceroici:2017:Forces}. This 2D imaging method is called Fast Orthogonal Row-Column Electronic Scanning (FORCES), and while it can create high quality images, it is 
limited to B-Mode imaging as there is no out-of-plane focusing. The B-Mode scan plane can be electronically steered, but 
constructing a volume at high resolution would take a very large amount of transmits. In another method we perform volumetric imaging with a receive aperture 
that mimics an ideal matrix probe, providing comparable volume quality to other existing RCA methods with the added benefit of being able to image beyond the 
shadow of the array \cite{Dahunsi:2025:HERCULES:Preprint}. This imaging method is called Hadamard Encoded Row-Column Ultrasonic Expansive Scanning (HERCULES), and it uses Hadamard 
encoding to create a virtual 2D receive aperture that can perform ideal spherical receive focusing. While HERCULES can make a volume at the same rate that 
FORCES can make a B-Mode slice, it is of lower quality, similar to other modern RCA imaging methods, due to HERCULES' lack of transmit focusing. This work 
introduces an iteration of HERCULES that can simultaneously provide transmit and receive focusing and obtain volumes with the quality of FORCES that have 
isotropic resolution. This imaging scheme is called Orthogonal Plane-Wave Transmit-Receive Isotropic-Focusing Micro-Ultrasound (OPTIMUS).

\section{Methods}

OPTIMUS relies on the use of the Hadamard-Encoded Read-Out (HERO) scheme to perform receive aperture encoding. This imaging method is described in full in 
\cite{Dahunsi:2025:HERCULES:Preprint}, but we will give a brief description here.

We can model a row-column array of $R$ rows and $C$ columns as $R \times C$ 'physical elements'. The signal received at row $r$ and column $c$ for some emission pattern 
is $s_{rc}(t)$. Let's say we receive an ultrasound signal on the columns of an electrostrictive array held at DC ground while applying DC biases on the rows for 
$E$ transmit events. If we construct a $R \times E$ matrix $H$ from the bias of each row $r$ for each transmit event $e$, we can express the signal received 
on column $c$ for each transmit event $e$, $g_c^{\{e\}}(t)$ as:

\begin{equation}
    g_c^{\{e\}}(t) = \sum_{r} H_r^{\{e\}}s_{rc}(t)  
\end{equation}

In matrix form, this is $\mathbf{g = Hs}$, and thus we can create an estimate for the receive signal data $\mathbf{\hat{s} = H^{-1}g}$. A HERO acquisition is where we 
take $H$ to be a $R \times R$ Hadamard matrix, and after $E = R$ transmit events we decode our channel data to obtain a signal dataset $\hat{s}_{rc}(t)$ that represents 
a virtual 2D receive aperture.

Performing a HERO acquisition will give us an effective receive aperture that matches a matrix array with full spherical focusing. However, our transmit aperture
is completely unfocused. To provide transmit focusing OPTIMUS will repeatedly perform HERO acquisitions with varying transmit delay profiles. As we are still 
using a row column array, this will only provide cylindrical transmit focusing, but combined with full spherical receive focusing this provides near isotropic
resolution.  Figure \ref{fig:imagingScheme} shows a summary of this imaging scheme.

\subsection{Simulation}

To assess the isotropy of OPTIMUS we simulated a field of scatterers using FieldII. Scatterers are placed at various positions in the transducer's
field of view and we measure the point spread function profiles and how they vary across space. Table \ref{table:simDieParams} a), b) \& d) lists the simulation 
parameters.

\begin{figure}
    \centering
    \includegraphics[width=0.9\columnwidth]{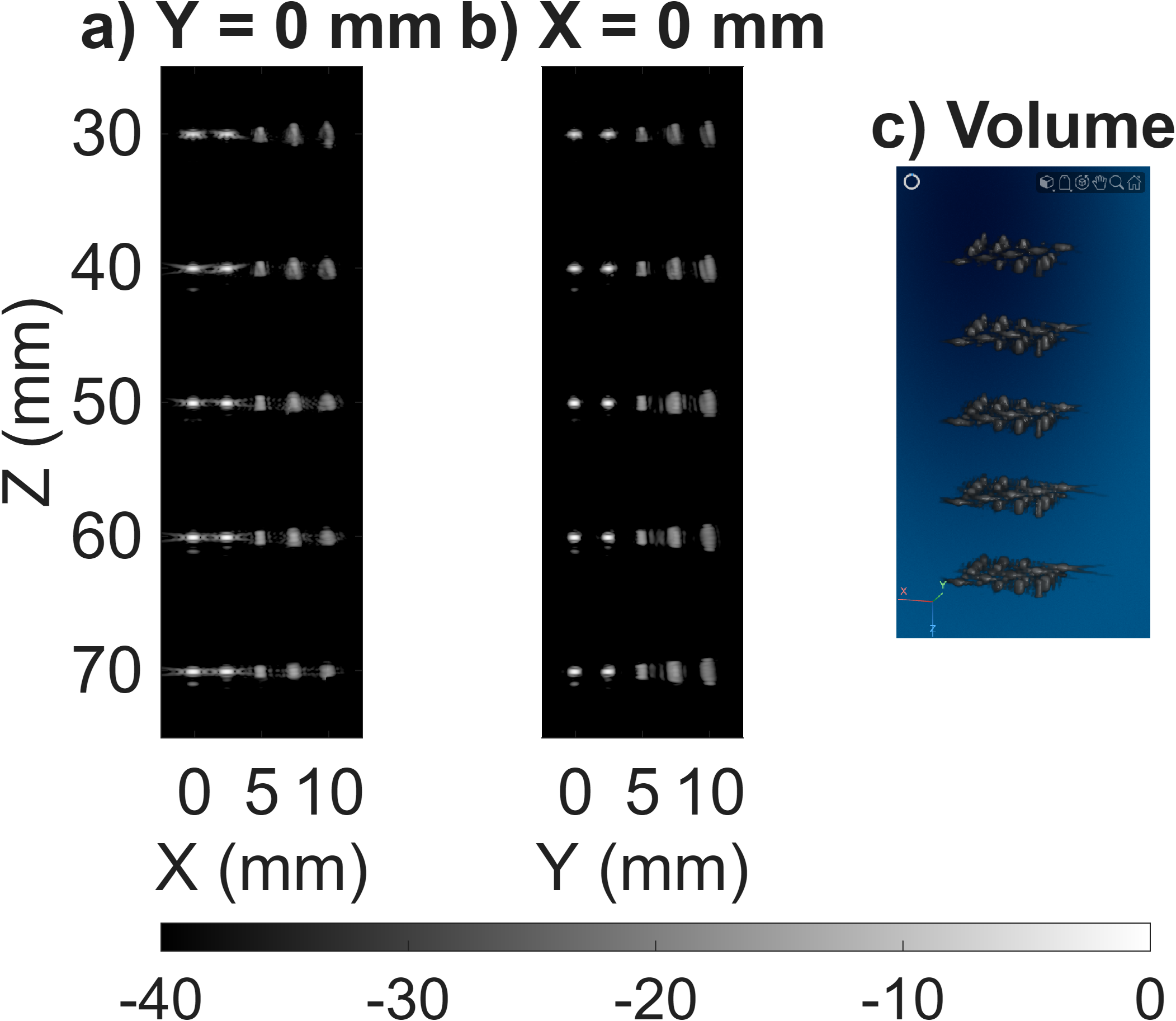}
    \caption{Simulations of a grid of point scatters imaged with OPTIMUS. a-b) are maximum intensity projections over a 5 mm slice, centered around a single 
    plane of points in the XZ, YZ \& XY planes, respectively. c) is a volume render of the 3D grid of scatters.}
    \label{fig:psfGrid}
\end{figure}

\subsection{Experiment}

\begin{figure*}
    \centering
    \includegraphics[width=0.9\linewidth]{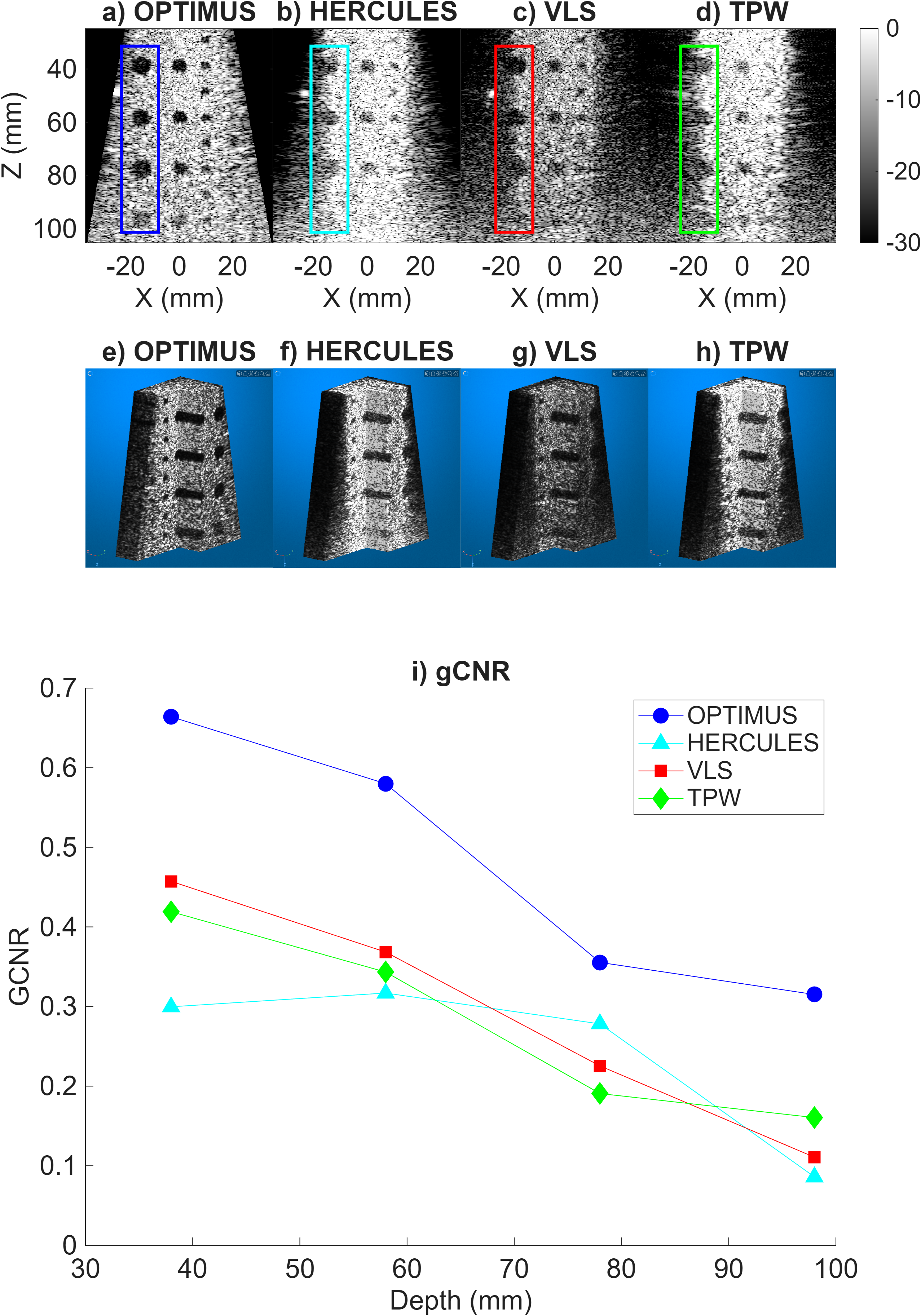}
    \caption{Experimental volume imaging of a commercial calibration phantom (CIRS ATS 539). a-d) show the center slice of the volumes, and e-h) show cut-outs.
    i) is a measurement of the gCNR of the largest cysts. The HERCULES, VLS \& TPW volumes were created using 128 emissions, while the OPTIMUS volume was 
    created using 1152 emissions. }
    \label{fig:cystComparison}
\end{figure*}

\begin{figure}
    \centering
    \includegraphics[width=0.6\columnwidth]{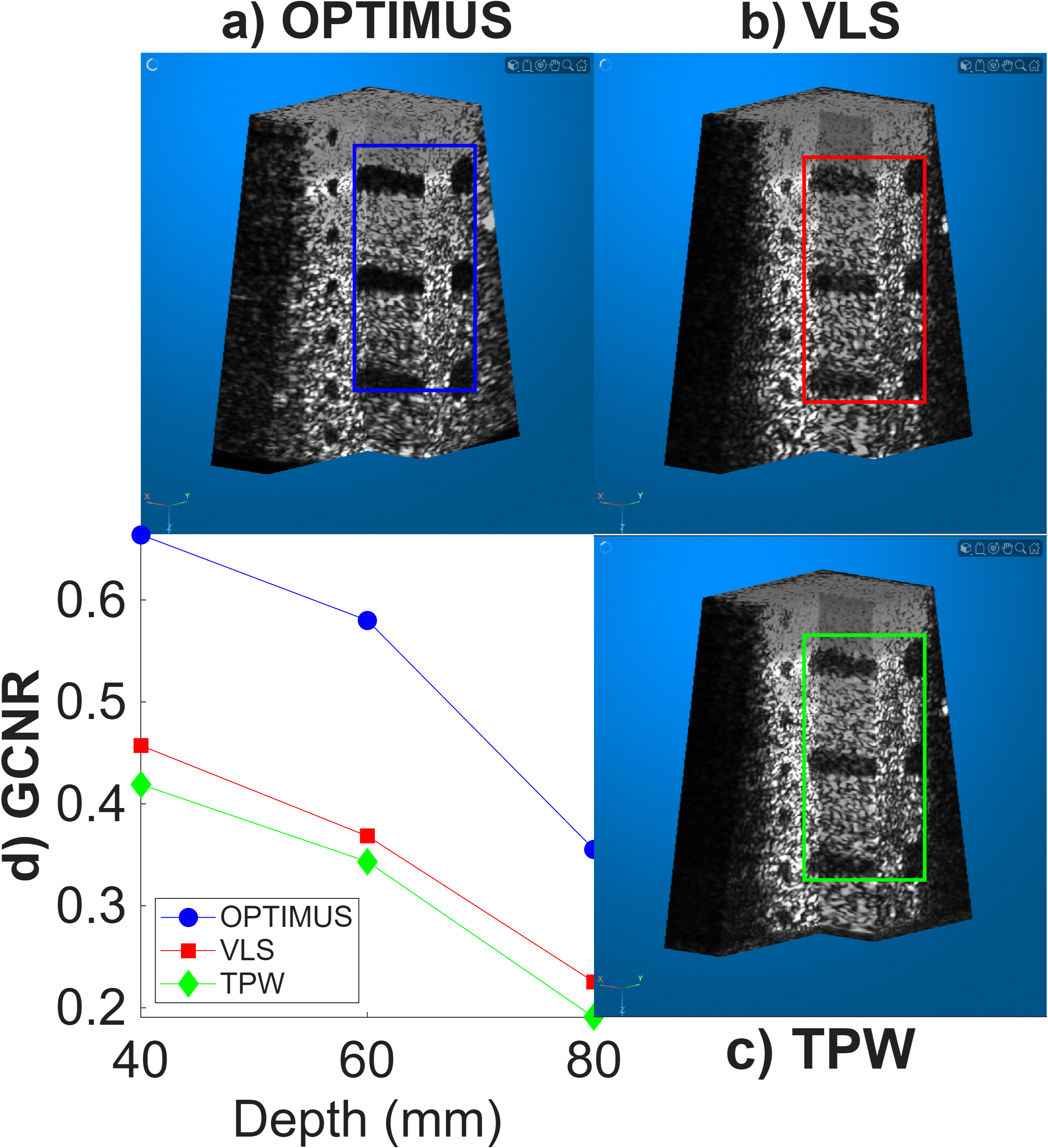}
    \caption{Experimental comparisons of OPTIMUS with standard RCA imaging methods using an equal number of emissions.  All 3 volume were created using 1152 
    emissions. The gCNR of the largest cysts are plotted in d).}
    \label{fig:transmitCount}
\end{figure}

We used a custom-fabricated 4-8 MHz 128$\times$128 TOBE array (CliniSonix, Edmonton, AB, Canada) \cite{Sampaleanu:2014:CMUT_TOBE_Ultrasound, 
Chee:2014:CMUT_TOBE_Photoacoustic} to implement OPTIMUS and image various commercial phantoms and ex-vivo samples. We used custom biasing electronics connected     
to a Vantage 256 High Frequency Ultrasound System (Verasonics, Kirkland, WA, USA) \cite{Ilkhechi:2023:BiasSwitching}. We couple DC signals from the biasing 
electronics and AC signals from the Vantage to drive the transducer. Figure \ref{fig:experimentalSetup} shows the experimental setup, and Table 
\ref{table:simDieParams} a), c) \& d) lists the experimental properties.

We imaged a commercial quality assurance phantom (ATS 539, CIRS Inc., Norfolk, VA, USA) to characterize the quality of our imaging scheme. We compared OPTIMUS 
with other volumetric imaging techniques such as HERCULES, Virtual Line Source (VLS), \& Tilted Plane-Wave Imaging \cite{Dahunsi:2025:HERCULES:Preprint, 
Rasmussen:2015:3DRCA1, Chen:2018:RCA3DPW}. We also imaged an ex-vivo pig fetus to demonstrate real tissue imaging capabilities. The fetus was immersed in a water 
bath for coupling. This imaging was performed using chirp excitations to compensate for reduced SNR caused by unfocused insonification.

\subsection{Processing}

After receiving data from the Vantage system, we apply a matched filter and demodulate our signal to baseband. For our encoded imaging schemes, HERCULES \& 
OPTIMUS, we need to perform a decoding step, where we use the inverse of the Hadamard matrix $H$ to recover $s_{rc}$. This provides us with the dataset of a 
virtual 2D receive aperture. At this point we can apply standard synthetic aperture delay-and-sum beamforming and obtain a volume.

\begin{table}[htbp]
    \centering
    \caption{Scatter Grid Resolutions}
    \label{tab:gridResolution}
    \begin{tabular}{lcc}
        \hline
        Metric & Normalized ($\mu m$ per $\lambda f\#$) \\ \hline\hline
        Lateral Resolution & $1.60 \pm 0.46$ \\ 
        Elevational Resolution & $1.42 \pm 0.39$ \\
        \hline
        Metric & Normalized ($\mu m$ per $\lambda f\#$) \\ \hline\hline
        Axial Resolution & $2.76 \pm 1.91$ \\
        \hline
    \end{tabular}
\end{table}

\section{Results}

\begin{figure*}
    \centerline{\includegraphics[width=\linewidth]{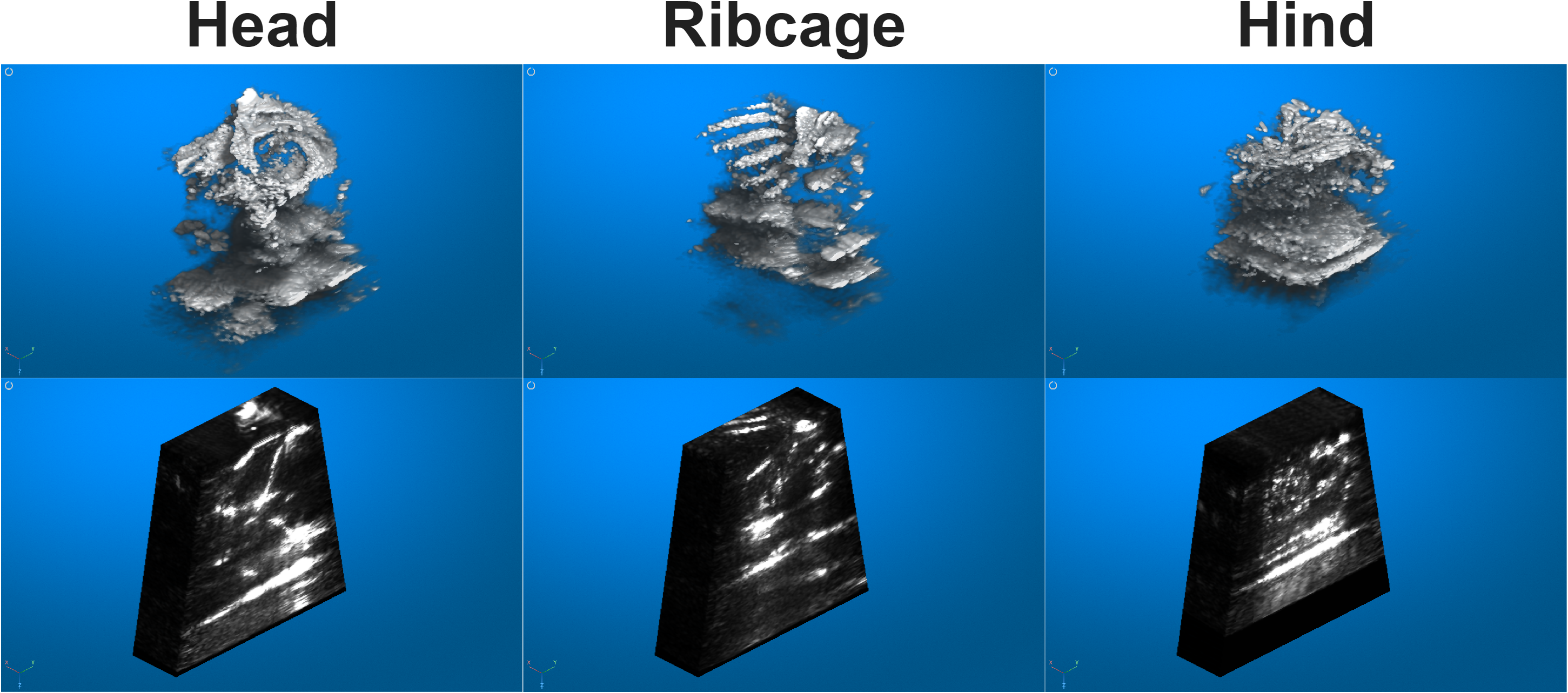}}
    \caption{Ex-vivo Fetal Pig. Fetus was imaged while immersed in a water bath to maximize coupling with the transducers. Using OPTIMUS we generated expansive
    volumes of the fetus at it's head, ribcage and hind legs. A maximum intensity projection is shown on the top figures, while a cut plane at center of the 
    volumes is shown on the bottom. Videos are provided in the supplementary materials.}
    \label{fig:fetalPig}
\end{figure*}

Figure \ref{fig:psfGrid} displays some sample point-spread-functions (PSFs) from our scatterer grid simulation. In Table \ref{tab:gridResolution} we plot 
resolution metrics as they vary across the medium. The average lateral and elevational resolution match the theoretical lateral resolution of 
$\text{FWHM} = 1.4F\# = 1.4*f/D$ where FWHM is the full-width-half-maximum value of the profile, $F\#$ is the f-number which is equivalent to the ratio of the 
focal length $f$ and the characteristic length of the aperture $D$. The lateral and elevational resolution does increase as points go deeper, but that is 
expected as the $F\#$ increases. The variation as x or y changes is relatively small. The axial resolution increases in variance away from the central axis of 
the transducer, potentially decreasing fidelity at the edges of the volume. Typical 3D imaging schemes using 2D arrays have PSFs that vary significantly across 
the array's field of view, but OPTIMUS is able to maintain relative consistency, nearly up to the edge of the shadow of the aperture.

Experimental characterization of OPTIMUS is shown in our phantom images in Figure \ref{fig:cystComparison}. It is compared with HERCULES, VLS, \& TPW. HERCULES, 
VLS \& TPW all show comparable results, but VLS \& TPW are unable to image beyond the shadow of the aperture. All three methods show ideal resolution along the 
receive direction, but display quality degradation in the transmitting direction. In contrast, OPTIMUS does not display any difference in the transmit direction. 
The figure also shows a quantitative comparison of the imaging method's contrast. OPTIMUS displays significantly higher gCNR than any of the other imaging 
methods. Videos of other imaging phantoms imaged with OPTIMUS are provided in the supplementary material.

To evaluate whether OPTIMUS' performance is purely a product of increased SNR due to a higher transmit count, we compared 
OPTIMUS to VLS \& TPW acquisitions consisting of an equal number of transmits. Figure \ref{fig:transmitCount} shows both visual and quantitative comparisons. The 
marginal improvements of VLS \& TPW diminish as the transmit count increases, and they are unable to achieve the same level of quality that OPTIMUS can. In a 
clinical scenario where this extended acquisition is not problematic, OPTIMUS provides significant advantages.

Figure \ref{fig:fetalPig} demonstrates in-vivo imaging of a fetal pig with OPTIMUS. Volume flythroughs are provided in the supplementary material. In these volumes
we can distinguish organs in various portions of the body. The large footprint of our array coupled with the wide FOV of OPTIMUS allow us to image a large 
portion of the fetus.

The main drawback of an OPTIMUS acquisition is the transmit count. The volumes in the previous figures are acquired at 9 angles, and when combined with decoding
on a $128\times128$ array requires 1152 transmit-receive events. At a 10 KHz base PRF we are limited to a 8.7Hz acquisition rate. This is sufficient for ex-vivo 
imaging, but potentially prohibitive for clinical applications where large motion is expected. However, these problems can potentially be mitigated with motion 
compensation between lower resolution volumes. These low resolution volumes could be acquired by using 128 emissions at a single angle, but a partial decoding 
technique could be used to reduce the requred emissions to as few as 16 \cite{Henry:2025:READIEMC2:Preprint}.

\section{Conclusion}

In this work we demonstrated a new imaging method that performs high quality isotropic volumetric imaging using the properties of bias-sensitive row column 
arrays. This method has superior resolution and contrast to other existing volumetric imaging methods, at the cost of a slow acquisition rate. This makes it a 
good candidate for imaging static features and generating structural images, such as visualizing the margins of an excised tumor. Motion 
compensation can be leveraged to mitigate the negative effects of compounding. Future work could include using an array with an acoustic lens to acquire full 
pyramidal volumes, or applying computer vision techniques to perform automatic segmentation of tissue structures in realtime.

\bibliographystyle{IEEEtran}
\bibliography{optimus}

\end{document}